\documentclass[twocolumn,showpacs,preprintnumbers,amsmath,amssymb,groupedaddress,prb]{revtex4}

\usepackage{graphicx}
\usepackage{dcolumn}
\usepackage{bm}
\usepackage{epsfig}

\def\e{\begin{equation}}
\def\f{\end{equation}}
\def\=#1{\overline{\overline #1}}

\def\-#1{{\bf #1}}
\def\_#1{{\bf #1}}
\def\o{\omega}
\def\v{\varepsilon}
\def\M{\mu}
\def\M0{\mu_0}

\def\.{\cdot}

\def\##1{{\bf#1\mit}}
\def\Re{{\rm Re\mit}}

\def\l#1{\label{eq:#1}}
\def\r#1{(\ref{eq:#1})}

\def\vec#1{{\bf #1}}

\begin{document}

\title{On the low-frequency spatial dispersion in wire media}

\author{C.R. Simovski}
\affiliation{Photonics and Optoinformatics Department, St. Petersburg State University of Information Technologies, Mechanics and Optics, Sablinskaya 14, 197101, St. Petersburg, Russia}
\email{belov@rain.ifmo.ru}

\author{P.A. Belov}
\affiliation{Photonics and Optoinformatics Department, St. Petersburg State University of Information Technologies, Mechanics and Optics, Sablinskaya 14, 197101, St. Petersburg, Russia}

\date{\today}

\begin{abstract}
The work is dedicated to the theoretic analysis of wire
media, i.e. lattices of perfectly conducting wires comprised of two or three
doubly periodic arrays of parallel wires which are orthogonal to one another. 
An analytical method based on local field approach is used.
The explicit dispersion equations are presented and studied. 
A possibility to introduce a dielectric permittivity is discussed. The theory is validated by comparison with the numerical data available in the literature.
\end{abstract}

\pacs{41.20.Jb, 
42.70.Qs, 
77.22.Ch, 
77.84.Lf 
}

\maketitle

\section{Introduction}

In the recent years the periodic metallic lattices have found many
applications both in optical and microwave ranges (see, for
example, in \cite{Sakoda} and \cite{Thevenpot}). However,
some fundamental problems have not been resolved yet, even for typical metallic
electromagnetic crystals. One of them is the problem of
low-frequency spatial dispersion in wire media (WM). The
low-frequency spatial dispersion of a simple wire medium (a doubly
periodic regular array of parallel wires) has been studied only
recently in \cite{WMPRB}. In the present paper
this theory is generalized for double and triple wire media. The study of spatial dispersion effects in 
the above mentioned variants of WM has been started in work \cite{Mario3d}. However, this  
study (based on the numerical approach) is far from being complete.  
Our theory significantly complements the results \cite{Mario3d}. It
is analytical one, and in order to validate it, a comparison to the results from \cite{Mario3d} is carried out.

\begin{figure}[htbp]
\centering \epsfig{file=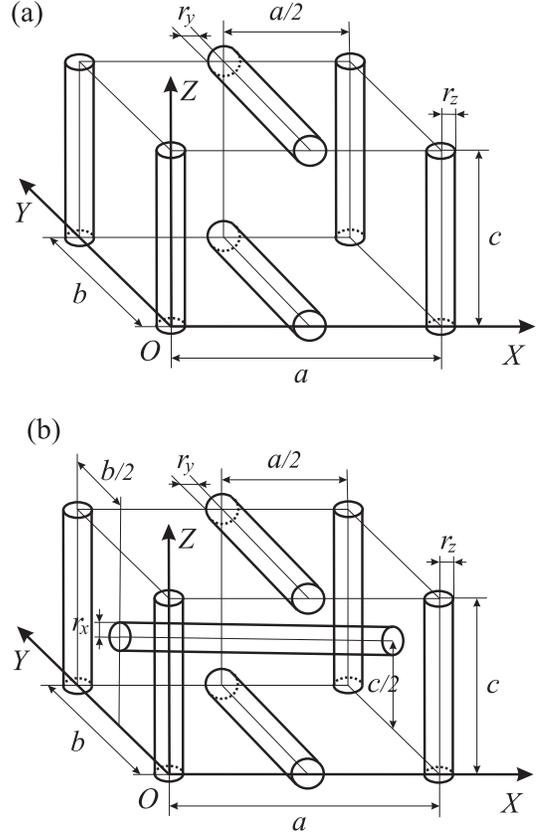, width=7cm} \caption{Unit cells of 
double wire medium (a) and triple wire medium (b).} \label{geom}
\end{figure}

The unit cells of lattices under study are shown in Fig. \ref{geom}. They are comprised of
two (2d or double  wire medium) or three (3d or triple wire medium) doubly periodic regular arrays of parallel infinite wires which are
orthogonal to one another.
The wires are assumed to be perfectly conducting. The host medium
is a uniform lossless dielectric with permittivity $\varepsilon_0$ and permeability $\mu_0$. We
denote the radii of wires directed along $x$,$y$ and $z$-axes as $r_x$, $r_y$ and $r_z$, respectively.
The periods of the lattice along $x$,$y$ and $z$-axes are denoted as $a$, $b$ and $c$, respectively. 
The lattices are spatially shifted with respect to each other by half period (see Fig. \ref{geom}).
The wires axes positions in the chosen coordinate system are determined by equations:
\begin{itemize}
\item the $x$-directed wires: $y=bn+b/2$ and $z=cl+c/2$, 
\item the $y$-directed wires: $x=am+a/2$ and $z=cl$, 
\item the $z$-directed wires: $x=am$ and $y=bn$, 
\end{itemize}
where $m$, $n$ and $l$ are integers.

In order to model an electromagnetic response of a wire we apply the local field approach. 
We assume that the wires diameters are small as compared to the wavelength. 
Thus, every wire can be described in terms of effective linear current referred to the wire axis.
The wire with radius $r_0$ oriented along a unit vector $\vec d$ ($|\vec d|=1$) can be characterized by a "polarizability" $\alpha$,
relating the complex amplitude $I$ of the induced current and the local electric field $\vec E^{\rm loc}$: 
\e
I=\alpha(r_0,k,\vec q\cdot \vec d) \vec E^{\rm loc}\cdot \vec d.
\l{alphadef}
\f
Here $k$ is the wave number of the host medium, and $\vec q\cdot \vec d=q_\parallel$ is the longitudinal component of the wave vector of propagating mode.  The following expression for $\alpha$ was obtained in \cite{WMJEWA}: 
$$
\alpha(r_0,k,q_\parallel)=
$$
\e
=\left[\frac{\eta(k^2-q_\parallel^2)}{4k}
\left(1-j\frac{2}{\pi}\left\{\log\frac{\sqrt{k^2-q_\parallel^2}\, r_0}{2}+\gamma\right\}\right)
\right]^{-1},
\l{alpha}
\f
where $\gamma\approx 0.5772$ is the Euler constant and $\eta=\sqrt{\mu_0/\varepsilon_0}$ is the wave impedance of the host medium.

Let the eigenmode under consideration have the wave vector $\vec q=(q_x,q_y,q_z)^T$.
Expressing the local field produced by all wires except the reference one through the current induced in the reference wire we obtain the dispersion equation. It  
relates the components of $\vec q$ with $k$ (i.e. with frequency $\omega$). Then we can 
introduce effective material parameters of the wire medium which fit this dispersion equation. In this paper we widely use the results obtained in our 
precedent papers \cite{WMPRB,WMJEWA} for a simple WM. Therefore, in the next section a very short overview of those works is presented. 

\section{Simple wire media}

\begin{figure}[h]
\centering \epsfig{file=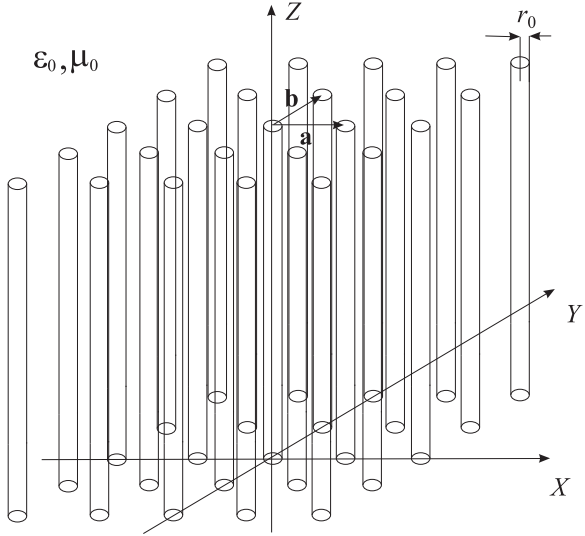, width=8cm} \caption{Simple wire media: a doubly periodic lattice of parallel ideally conducting thin wires.} \label{1d}
\end{figure}

The geometry of a simple wire medium comprising the $z-$directed wires is shown in Fig. \ref{1d}.

The currents in the wire with numbers $(m,n)$ (counted along $x$ and $y$ axes, respectively) are related with current $I$ induced in the reference (zeroth) wire through the wave vector $\vec q$:
\e
I_{m,n}=Ie^{j(q_xam+q_ybn)}.
\f
Since the field re-radiated by the wire $(m,n)$ is proportional to $I_{m,n}$, the local electric field acting to the zeroth wire can be expressed in terms of the so-called {\it dynamic interaction constant} $C$:
\e
E_z^{\rm loc}=C(k,q_x,q_y,q_z,a,b)I,
\l{Cdef}
\f 
where (see in \cite{WMPRB,WMJEWA}):
$$
C(k,q_x,q_y,q_z,a,b)=-\frac{\eta(k^2-q_z^2)}{4k}\times 
$$
\e\sum\limits_{(m,n)\ne (0,0)}
\left[H_0^{(2)}\left(\sqrt{k^2-q_z^2}R_{m,n}\right)e^{-j(q_xam+q_ybn)}\right],
\l{Cgen}
\f 
$R_{m,n}=\sqrt{(am)^2+(bn)^2}$ and all $(m,n)$ except $m=n=0$ are summed up. The expression \r{Cgen} can be rewritten in the following form 
\cite{WMJEWA}:
$$
C(k,q_x,q_y,q_z,a,b)=
\frac{\eta(k^2-q_z^2)}{2jkb}\left[
\frac{1}{k_x^{(0)}}\frac{\sin k_x^{(0)}a}{\cos k_x^{(0)}a-\cos q_xa }\right.
$$
$$
+\sum\limits_{n\ne 0}
\left(\frac{1}{k_x^{(n)}}{\frac{\sin k_x^{(n)}a}{\cos k_x^{(n)}a-\cos q_xa }}-\frac{b}{2\pi |n|}\right)
$$
\e
\left.
+\frac{b}{\pi}\left(\log{\frac{\sqrt{k^2-q_z^2}b}{4\pi}}+\gamma\right)+j\frac{b}{2}\right].
\l{C}
\f
Where
\e
k_x^{(n)}=-j\sqrt{\left(q_y+\frac{2\pi n}{b}\right)^2+q_z^2-k^2}, 
\l{kf}
\f
and we choose $\Re\{\sqrt{()}\}>0$. Those formulae physically correspond to the representation of the WM as a set of parallel grids (of $z-$directed wires) located parallel to one another with period $a$ along the $x-$axis (see also 
Fig. \ref{2d} for the case of 2d WM). Every grid radiates the spectrum of Floquet harmonics with wave vectors $(k_x^{(n)},q_y+2\pi n/b,q_z)$. The series with summation over $n$ in the right side of \r{C} describes the contribution of the high-order Floquet modes into electromagnetic interaction of those grids. 
The dispersion equation follows from \r{alphadef} and \r{Cdef}:
\e
\left[\alpha^{-1}(r_0,k,q_z)-C(k,q_x,q_y,q_z,a,b)\right]I=0
\l{dispgen}
\f

\begin{figure}[h]
\centering 
\epsfig{file=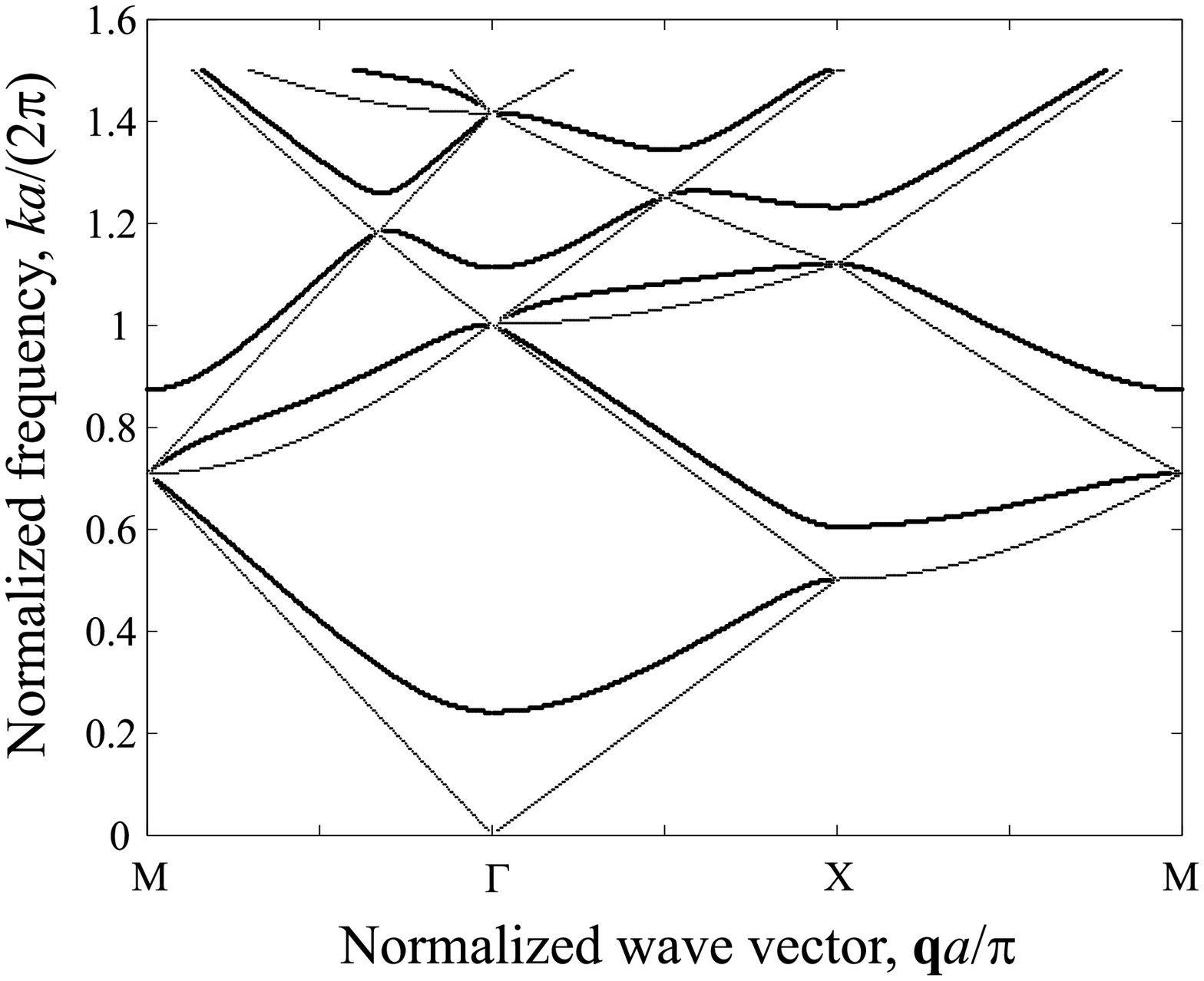, width=8cm}
\caption{Dispersion curves of wire media with filling ratio $f=\pi r^2_0/a^2=0.001$ (square lattice). Thin lines -- ordinary waves, thick lines -- extraordinary waves. TLM are not shown.}
\label{d1d}
\end{figure}

Taking into account expressions \r{alpha} and \r{C}
one can rewrite \r{dispgen} in the following form:
$$
\left(k^2-q_z^2\right)\left[\frac{1}{\pi}\log{\frac{b}{2\pi r_0}}+
\frac{1}{bk_x^{(0)}}
\frac{\sin k_x^{(0)}a}{\cos k_x^{(0)}a-\cos q_xa }+\right.
$$
\e
\left.\sum\limits_{n\ne 0}
\left( \frac{1}{bk_x^{(n)}}
\frac{\sin k_x^{(n)}a}{\cos k_x^{(n)}a-\cos q_xa }-\frac{1}{2\pi |n|}\right)\right]I=0.
\l{dis1}
\f
This equation has three types of solutions:
\begin{enumerate}
\item Ordinary waves, in case when $I=0$ in \r{dis1}. They have no electric field component along wires ($E_z=0$) and propagate without interaction with the lattice. Their dispersion plot corresponds to the host medium and is shown in Fig. \ref{d1d} by thin lines.
\item Extraordinary waves, in case when the expression in square brackets in \r{dis1} equals to zero. They correspond to the nonzero currents $I\ne 0$ and have the nonzero longitudinal component of electric field $E_z\ne 0$. Their dispersion properties are described in details in \cite{WMJEWA}; their dispersion curves are presented in Fig. \ref{d1d} by thick lines. 
\item Transmission-Line Modes (TLM), in case when $(k^2-q_z^2)=0$ in \r{dis1}. Those waves propagate along the wires, they are TEM waves ($E_z=0$), but $I\ne 0$. Their dispersion equation $q_z^2=k^2$ has no restriction for components $q_x,q_y$, and the phase shift of the currents in the adjacent wires can be arbitrary \cite{WMPRB}.
\end{enumerate}

Under the quasi-static limit $ka\ll 2\pi$ and $|q|a\ll 2\pi$, the dispersion equation for extraordinary waves transforms to 
\e
q^2=q_x^2+q_y^2+q_z^2=k^2-k_0^2,
\l{qq}
\f
where the following notations are used:
\e
k_0^2=\frac{2\pi/s^2}{\displaystyle\log\frac{s}{2\pi r_0}+F(r)},
\l{k0}
\f
$s=\sqrt{ab}$, $r=a/b$ and $F(r)=F(1/r)$ is given by
\e
F(r)= -\frac{1}{2}\log r+ \sum\limits_{n=1}^{+\infty}
\left(\displaystyle \frac{\mbox{coth}(\pi n
r)-1}{n}\right)+\frac{\pi r}{6} \l{Ffun}. \f 
Parameter $k_0$ corresponds to the effective plasma frequency of the lattice $\o_0=k_0/\sqrt{\varepsilon_0\mu_0}$. 
For square lattices $a=b$ one has $F(1)=0.5275$.
Comparing \r{qq} with the well-known dispersion equation of uniaxial dielectrics we obtain an effective relative permittivity $\=\epsilon$ of 1d WM in the following form:
\e
\=\varepsilon=
\varepsilon\_z_0\_z_0+\_x_0\_x_0+\_y_0\_y_0,\l{uni}
\f 
\e
\varepsilon(k,q_z)=
1-\frac{k_0^2}{k^2-q_z^2}. \l{eps} \f  
The dependence of dielectric permittivity on $q_z$ given by \r{eps} does not disappear untill frequency becomes zero. That fact means the low-frequency spatial dispersion of wire media.
There is no low-frequency spatial dispersion for the extraordinary waves in the only case when the wave 
propagates across the wires ($q_z=0$). At low frequencies the propagation of those waves can be described in terms of plasma-like permittivity $\varepsilon=1-k_0^2/k^2$ (see also \cite{Brown}). Relative to those waves the wire medium behaves as a cold non-magnetized plasma
(a continuous dielectric medium). In other propagation directions the wire medium behaves differently. In \cite{WMPRB} we discuss the importance of the low-frequency spatial dispersion in 1d wire media. Below we theoretically show this phenomenon in 2d and 3d WM.

\section{Double wire media}

\begin{figure}[h]
\centering \epsfig{file=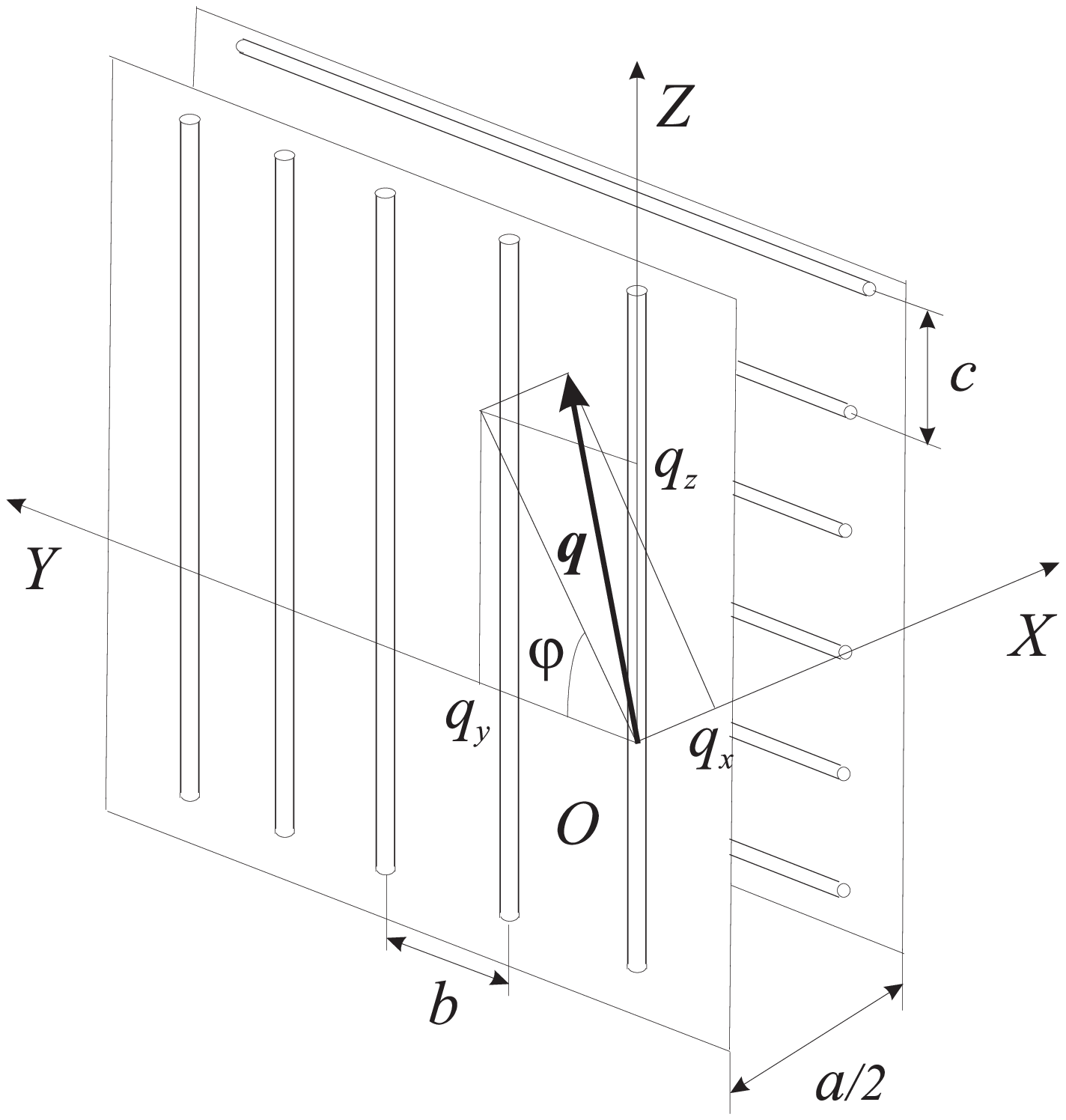, width=6cm} \caption{The structure of a
double wire medium is represented as a set of planar wire
grids.} \label{2d}
\end{figure}

Now, let us consider a double wire medium which is comprised by $y-$directed and $z-$directed wires. 
It is shown in Fig. \ref{2d} as a set of parallel grids located along the $x-$axis.  
Below the local field approach is going to be applied taking the same approximation as it was done in
our work \cite{APSWSRR} (where we have studied a doubly negative meta-material in a similar way). Thus, the approximation is following: 
the electromagnetic field produced by a single grid of wires at the distance from the grid $a/2$
is considered as a field of a sheet of the average current $\vec J$. That approximation 
is accurate enough under the condition when the wavelength in the matrix is large as compared to the grid periods ($kb\ll 2\pi$ and $kc\ll 2\pi$) and period $a$ is not smaller than periods $b$, $c$. In this case the $y$-oriented grids interact with $z$-oriented grids by the fundamental Floquet harmonic. Other harmonics are evanescent and their contribution into this cross-polarized interaction is negligible.

We can express the $(m,n)$-numbered $z-$directed current through the reference (zeroth) $z-$directed current $I_z$:
\e
I_z^{(m,n)}(z)=I_ze^{j(q_xam+q_ybn+q_zz)}.
\f
The same rule holds for $y-$directed currents:
\e
I_y^{(m,l)}(z)=I_ye^{j(q_xam+q_zcl+q_yy)}.
\f

The currents $I_z$ and $I_y$ are related to the local electric fields acting on the $z-$ and $y-$directed 
reference wires $E_z^{\rm loc}$ and $E_y^{\rm loc}$ through polarizabilities $\alpha_{z,y}$:
\e
I_z=\alpha_zE_z^{\rm loc}, \quad I_y=\alpha_yE_y^{\rm loc},
\quad \alpha_{y,z}\equiv\alpha(r_{y,z},k,q_{y,z}),
\l{al2}
\f
which fit \r{alpha}. 
Both $E_z^{\rm loc}$ and $E_y^{\rm loc}$ contain contributions of $z-$ and $y-$arrays:
\e
E_z^{\rm loc}=E_z^{(z)}+E_z^{(y)}, \quad  E_y^{\rm loc}=E_y^{(y)}+E_y^{(z)}. 
\f

Co-polarized terms $E_z^{(z)}$ and $E_y^{(y)}$ could be expressed through $I_z$ and $I_y$ applying \r{Cdef}:
\e
E_z^{(z)}=C_{zz}I_z, \quad C_{zz}\equiv C(k,q_x,q_y,q_z,a,b),
\l{c2z}
\f
\e 
E_y^{(y)}=C_{yy}I_y, \quad C_{yy}\equiv C(k,q_x,-q_z,q_y,a,c).
\l{c2y}
\f

The cross-components $E_z^{(y)}$ and $E_y^{(z)}$ could be expressed through $I_y$ and $I_z$, respectively:
\e
E_z^{(y)}=C_{zy}I_y, \quad E_y^{(y)}=C_{yz}I_z.
\l{c2yz}
\f
The cross-polarized interaction factors $C_{yz,zy}$ are evaluated below. 
Substituting equations \r{c2z}, \r{c2y} and \r{c2yz} into \r{al2} we obtain a system of equations
\e
\left\{
\begin{array}{lcl}
(C_{yy}-\alpha^{-1}_y)I_y+C_{yz}I_z=0\\
C_{zy}I_y+(C_{zz}-\alpha^{-1}_z)I_z=0
\end{array}
\right.
\l{sys2d}
\f
First of all, it should be noticed that the solution of \r{sys2d} when $I_{y,z}=0$ corresponds to the ordinary waves with polarization along the $x-$axis, propagating in the plane $(y-z)$. The dispersion equation for such waves is $q_y^2+q_z^2=k^2$, $q_x=0$. Setting the determinant of \r{sys2d} equal to zero we obtain a dispersion equation for the extraordinary waves ($I_{y,z}\ne 0$):
\e
(C_{yy}-\alpha^{-1}_y)(C_{zz}-\alpha^{-1}_z)-C_{yz}C_{zy}=0.
\l{dis2d}
\f  
It should be noticed that the expressions in brackets in \r{dis2d} are exactly the dispersion equations for the simple wire media
(from $y$-wires and $z-$wires, respectively).

In \r{dis2d} coefficients $C_{yy,zz}$ are defined from \r{C}, \r{c2z}, \r{c2y}. Now, let us calculate coefficients $C_{zy}$ and $C_{yz}$ using the approximation of current sheets which has been mentioned above. The $z-$component of the electric field produced by a sheet with surface current
$\vec J_{y}(y,z)=\vec y_0(I_y/c)e^{j(q_yy+q_zz)}$ at the arbitrary distance $x$ from the grid
can be expressed by the formula \cite{Lindell}: \e
E_{y,z}(x,y,z)=\frac{\eta q_{y}q_{z}}{2k_x k}
 J_{z,y}(y,z)e^{-jk_x|x|},
 \l{kappa}
 \f
 $$
 k_{x}\equiv-j\sqrt{q_y^2+q_z^2-k^2}.
$$
Summing up \r{kappa} over all $m-$numbered layers we can write:
\e
C_{zy}=\frac{\eta q_{y}q_{z}}{2k_x kc}\sum\limits_{m=-\infty}^{+\infty}e^{-jq_xam}e^{-j|m-\frac{1}{2}|k_x a}. \l{c}
\f
The summation result is following: 
\e 
C_{zy}={\eta q_{y}q_{z}\over k_x kc}{je^{-jq_xa/2}\cos{(q_xa/2)}\sin {(k_xa/2)}\over \cos q_xa-\cos k_x a}.
\l{czy}\f
Taking into account the phase shift between $z$ and $y$ grids we obtain
$$ 
C_{yz}=e^{jq_xa}C_{zy}c/b
$$
\e
={\eta q_{y}q_{z}\over k_x kb}{je^{jq_xa/2}\cos{(q_xa/2)}\sin {(k_xa/2)}\over \cos q_xa-\cos k_x a}.
\l{cyz}
\f
Substituting \r{alpha}, \r{C}, \r{czy} and \r{cyz} into \r{dis2d} we derive an explicit dispersion equation:
\e
\left(k^2-q_y^2\right)\left[\frac{1}{\pi}\log{\frac{c}{2\pi r_y}}+
\frac{1}{ck_x}
\frac{\sin k_xa}{\cos k_xa-\cos q_xa }+\right.
\l{dis2dbig}
\f
$$
\left.\sum\limits_{n\ne 0}
\left( \frac{1}{c\beta_y^{(n)}}
\frac{\sin \beta_y^{(n)}a}{\cos \beta_y^{(n)}a-\cos q_xa }-\frac{1}{2\pi |n|}\right)\right]\times
$$
$$
\left(k^2-q_z^2\right)\left[\frac{1}{\pi}\log{\frac{b}{2\pi r_z}}+
\frac{1}{bk_x}
\frac{\sin k_xa}{\cos k_xa-\cos q_xa }+\right.
$$
$$
\left.\sum\limits_{n\ne 0}
\left( \frac{1}{b\beta_z^{(n)}}
\frac{\sin \beta_z^{(n)}a}{\cos \beta_z^{(n)}a-\cos q_xa }-\frac{1}{2\pi |n|}\right)\right]=
$$
$$
{4q^2_{y}q^2_{z}\over k_x^2 bc} \left({\cos{(q_xa/2)}\sin {(k_xa/2)}\over \cos q_xa-\cos k_x a}\right)^2,
$$
where
$$
\beta_z^{(n)}=-j\sqrt{\left(q_y+\frac{2\pi n}{b}\right)^2+q_z^2-k^2}, 
$$
$$
\beta_y^{(n)}=-j\sqrt{\left(q_z+\frac{2\pi n}{c}\right)^2+q_y^2-k^2}.
$$
The signs of all square roots are chosen so that $\Re\{\sqrt{()}\}>0$.
The dispersion equation \r{dis2dbig} cannot be simplified (except the quasi-static limit) even in a special case when $r_y=r_z$ and $a=b=c$. In fact,  
the perfect square can be obtained in the left side of \r{dis2dbig}
if one neglects the contribution of high-order Floquet harmonics expressed by $n-$series.
However, this approximation leads to the wrong results  for the shape of isofrequencies. Therefore we do not use it.  

The preliminary analysis of \r{dis2dbig} reveals some special solutions. 
There are two solutions which correspond to TLM: the first is $q_y=k,\ q_{z}=0$, $q_x$ is arbitrary; the second one is $q_z=k,\ q_{y}=0$, $q_x$ is arbitrary. Those waves propagate either along the $y-$wires (when the electric field averaged over the lattice unit cell is polarized along $z$) or along the $z-$wires (when the averaged electric field is polarized along $y$). They are TEM waves as well as TLM in simple WM \cite{WMPRB}. The component $q_x$ is a free parameter for TLM and
plays a role of a phase shift between the currents in the adjacent grids of wires \cite{WMPRB}. 
At the first sight, it seems strange that the electric field with non-zero $z-$component
can propagate along $y$ {\it across} the $z-$directed wires below the "plasma" frequency (which is the cut-off frequency for such waves in 1d WM). However, it is possible.  When the TLM propagates along the $y-$wires, all grids of $z-$wires are excited, however the superposition of their fields exactly vanish in the planes $x=am+a/2$, where the grids of $y-$wires are located. This result can be easily obtained analytically for arbitrary non-zero $q_x$ and $q_t=k$. Due to the same reason it is also possible for $y-$polarized TLM to propagate along $z$.   

When $q_x=\pi/a$ or $q_yq_z=0$ the right side of \r{dis2dbig} equals to zero and the equation splits into two separate equations similar to \r{dis1} and describing the extraordinary waves in two simple WM. For $q_y=0$ (or $q_z=0$)
the absence of the interaction between two simple WM is trivial since the propagation holds in the plane $(x-z)$ (or $(x-y)$) and the electric field is polarized orthogonally to $y-$directed wires (or to $z-$wires). However, the interaction between two 1d WM is also absent when $q_x=\pi/a$. At low frequencies $ka<1$ the equation $q_x=\pi/a$ corresponds to the excitation of TLM {\it in both y- and z-arrays} with polarization directions alternating along $x$. The existence of this kind of TLM (which does not transport energy at all) is specific for 2d WM. 

More detailed study of \r{dis2dbig} requires numerical calculations and their results are presented below.
    
\section{Triple wire media}
Analysis of triple wire media 
can be carried out in the same way as it was described above.
Similarly to \r{sys2d} we obtain:
\e
\left\{
\begin{array}{lcl}
(C_{xx}-\alpha^{-1}_x)I_y+C_{xy}I_y+C_{xz}I_z=0\\
C_{yx}I_x+(C_{yy}-\alpha^{-1}_y)I_y+C_{yz}I_z=0\\
C_{zx}I_x+C_{zy}I_y+(C_{zz}-\alpha^{-1}_z)I_z=0
\end{array}
\right.
\l{sys3d}
\f
where $C_{yy,zz,yz}$ are determined by \r{c2z},\r{c2y},\r{czy},\r{cyz} and \r{C} and  
\e
C_{xx}=C(k,q_y,q_z,q_x,d_y,d_z).
\l{gad}\f
Here $\alpha_i$ is denoted as $\alpha_i=\alpha(r_i,k,q_i)$ and the subscript $i$ means the Cartesian components $(x,y,z)$.  
Other interaction factors are as follows:
\e 
C_{xy}=j{\eta q_{x}q_{y}\over k_z ka}{\cos{(q_zc/2)}\sin {(k_zc/2)}\over \cos q_zc-\cos k_z c}e^{jq_zc/2},
\l{gadstvo}\f
\e 
C_{xz}=j{\eta q_{x}q_{z}\over k_y ka}{\cos{(q_yb/2)}\sin {(k_yb/2)}\over \cos q_yb-\cos k_yb}e^{jq_yb/2},
\f
\e 
C_{yx}=j{\eta q_{x}q_{y}\over k_z kb}{\cos{(q_zc/2)}\sin {(k_zc/2)}\over \cos q_zc-\cos k_z c}e^{-jq_zc/2},
\f
\e 
C_{zx}=j{\eta q_{x}q_{z}\over k_y kc}{\cos{(q_yb/2)}\sin {(k_yb/2)}\over \cos q_yb-\cos k_yb}e^{-jq_yb/2},
\l{gad1}\f
$$
k_y\equiv -j\sqrt{q_x^2+q_z^2-k^2}, \quad k_z\equiv -j\sqrt{q_x^2+q_y^2-k^2}.
$$
There are no ordinary waves in that medium since there are no vectors orthogonal to all wires simultaneously.
Determinant of \r{sys3d} gives the dispersion equation:
\e
(C_{xx}-\alpha^{-1}_x)(C_{yy}-\alpha^{-1}_y)(C_{zz}-\alpha^{-1}_z)
\l{dis3d}
\f
$$
-(C_{xx}-\alpha^{-1}_x) C_{yz}C_{zy}
-(C_{yy}-\alpha^{-1}_y) C_{xz}C_{zx}
$$
$$
-(C_{zz}-\alpha^{-1}_z) C_{xy}C_{yx}
+C_{xy}C_{yz}C_{zx}+C_{xz}C_{zy}C_{yx}=0.
$$

The dispersion equation \r{dis3d} with substitutions \r{c2z},\r{c2y},\r{gad}--\r{gad1} is the final result for the triple wire media.
The explicit equation is cumbersome and cannot be simplified in the general propagation case. However, in the special case when $q_x=0$ all cross-polarized interaction terms \r{gadstvo}--\r{gad1} vanish, and the system 
\r{sys3d} splits into two separate sets: the first one is the dispersion equation of the 1d WM $C_{xx}=\alpha^{-1}_x$, the second one is the system \r{sys2d}. The first case corresponds to the extraordinary waves propagating normally to the $x$-wires without interaction with $y-$ and $z-$wires. There is no spatial dispersion for those waves (see above). The second case corresponds to the in-plane propagation in 2d WM, which will be studied below. In the present paper we do not consider the general case of the wave propagation in 3d WM. 

\section{Dispersion diagrams and isofrequencies of a double wire medium}

\begin{figure}[h]
\centering 
\epsfig{file=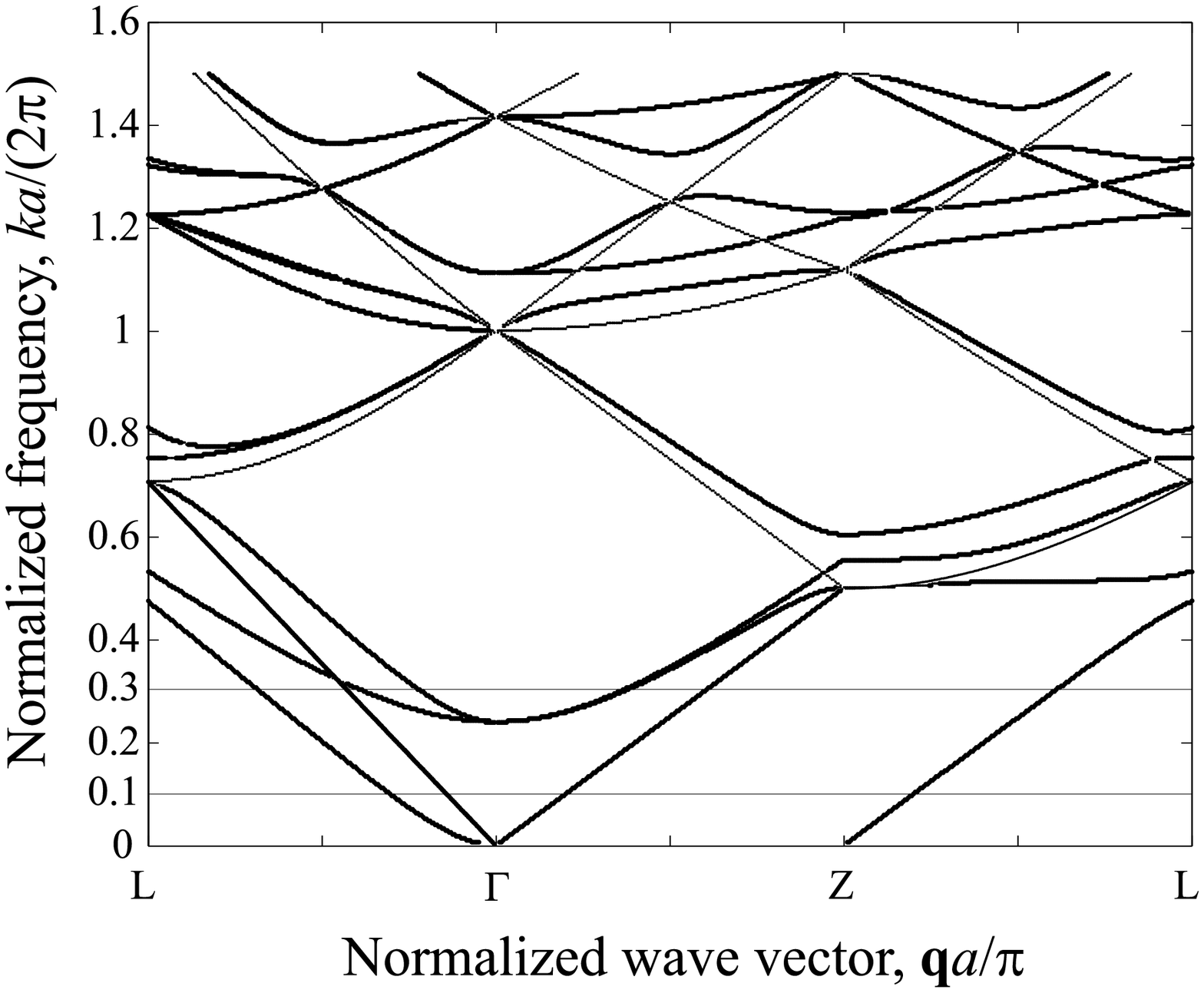, width=8cm}
\caption{Dispersion diagram of a double wire media with filling ratio $f=2\pi r^2_0/a^2=0.002$ (cubical cell and equal radii). Thin lines -- modes of the host medium (singular points of equation \r{dis2dbig}), thick lines -- modes of the 2d WM.}
\label{disyz}
\end{figure}

The dispersion diagram of a double WM for the in-plane propagation ($q_x=0$) of the extraordinary waves 
obtained by numerical solution of \r{dis2dbig} is shown in Fig. \ref{disyz}.
The chosen parameters of the wire lattice are $a=b=c$, $r_y=r_z$. The filling ratio is $f=2\pi r_y^2/a^2=0.002$.  
We use notations $\Gamma=(0,0,0)^T$, $Z=(0,0,\pi/c)^T$, and $L=(0,\pi/b,\pi/c)$
for the central point, the $z-$bound point and the corner point of the fundamental Brillouin zone, respectively.

One can notice the significant difference between Fig. \ref{disyz} and the dispersion diagram of a simple wire medium (see Fig. \ref{d1d}). In the Fig. \ref{disyz} one can see
within the interval $L-\Gamma$ two extraordinary modes which do not vanish at low frequencies $k< k_0$ and are not TLM. In simple WM the waves with nonzero longitudinal (with respect to the wires) component of the electric field cannot propagate at low frequencies since the phase shifts between the adjacent wires are small and the re-radiation of parallel wires suppresses the wave. In 2d WM it becomes possible due to the electromagnetic interaction of the two orthogonal wire arrays. This is the result of the cross-polarized interaction of wire arrays. There are terms in $C_{yz}$ that cancel out the terms in $C_{yy}$ and $C_{zz}$ which are responsible for the suppression of the waves propagating obliquely in simple WM at low frequencies. 

\begin{figure}[h]
\centering 
\epsfig{file=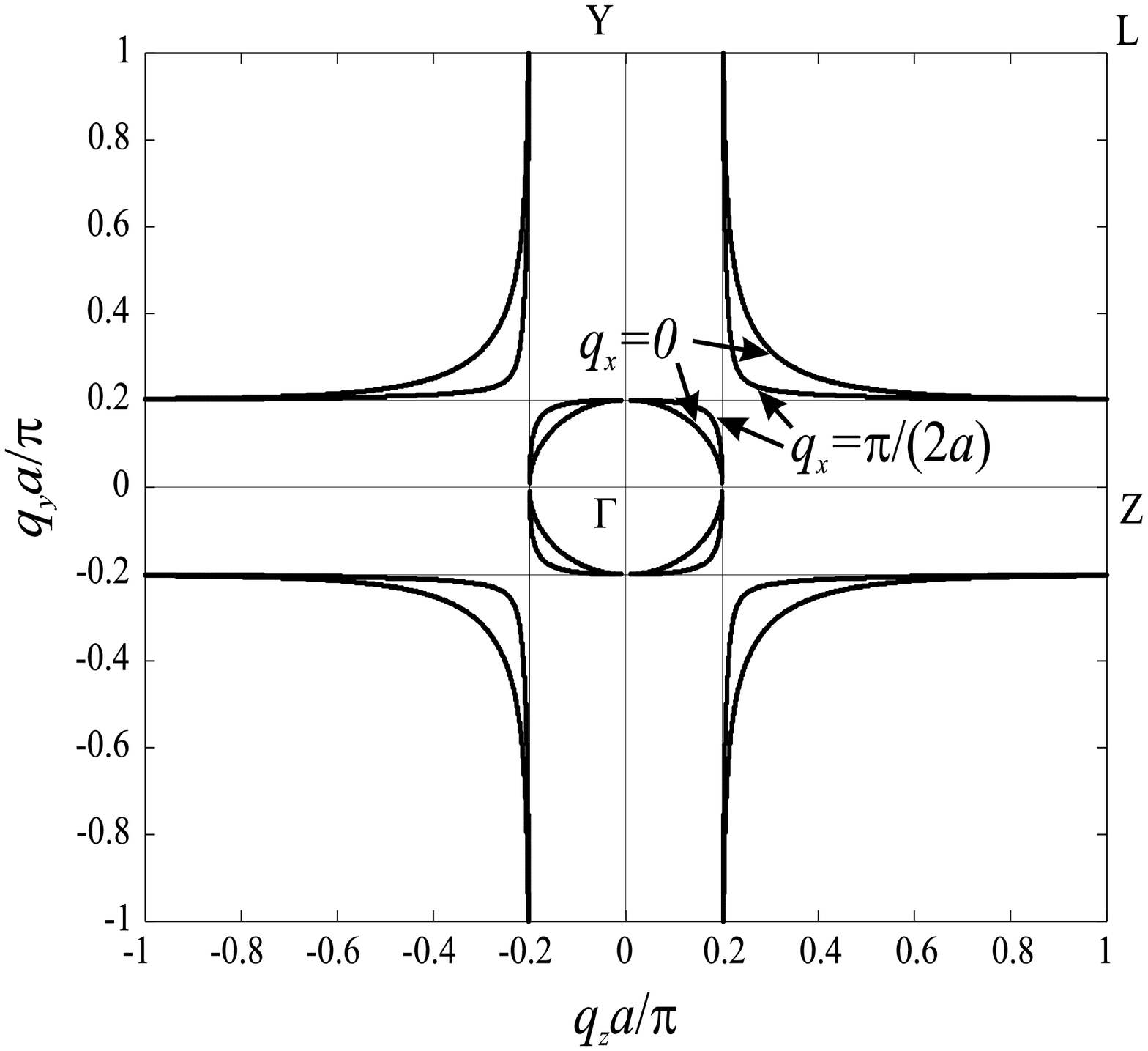, width=8cm}
\caption{Isofrequency contours for double wire media at $ka/(2\pi)=0.1$. Two cases $q_x=0$ and $q_x=\pi/(2a)$ are presented.}
\label{iso1}
\end{figure} 

\begin{figure}[h]
\centering 
\epsfig{file=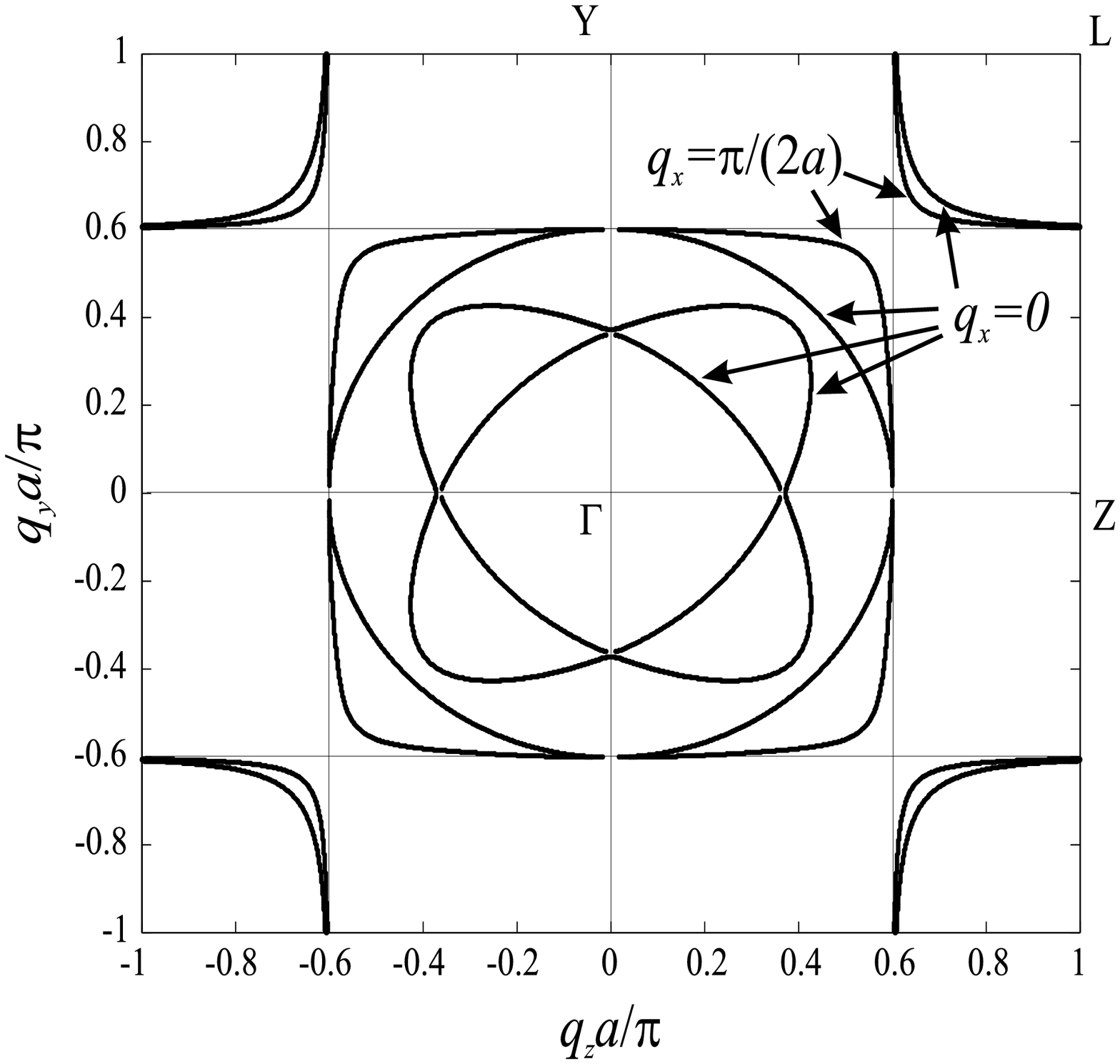, width=8cm}
\caption{Isofrequency contours for double wire media at $ka/(2\pi)=0.3$. Two cases $q_x=0$ and $q_x=\pi/(2a)$ are presented.}
\label{iso2}
\end{figure}

The horizontal lines $ka/(2\pi)=0.1$ and $ka/(2\pi)=0.3$ in Fig. \ref{disyz} correspond to isofrequency contours
presented in Fig. \ref{iso1} and \ref{iso2}, respectively.
The isofrequency contour located around the $L$ point is very unusual (close to the hyperbolic one).
In Fig. \ref{iso1} one can see that the contours of isofrequencies are rather close to four asymptotes $q_{y,z}=\pm k$. 
In spite of the rather low frequency as compared to $\o_0$, the isofrequency contour located around the $\Gamma$ point ($q=0$) basically differs from the isofrequency of an isotropic dielectric (a circle).  Only in a special case of the in-plane propagation the isofrequency centered at the $\Gamma$ point has practically circular shape and the phase velocity of this mode coincides with that of the host medium. When 
$q_x\ne 0$ the shape of this isofrequency becomes super-quadric and modes with hyperbolic isofrequency tend to the same asymptotes $q_{y,z}=\pm k$.   
When $q_x=\pi/a$ the isofrequencies coincide with the asymptotes exactly. This case corresponds to TLM discussed above (which do not transport energy).
The plot in Fig. \ref{iso1} indicates the possibility of the two refracted waves (both extraordinary waves) for the rather large sheer of incidence angles. This effect keeps at the quasi-static limit. 

The 2d WM with proper orientation of wires with respect to the medium interface can possess the low-frequency negative refraction. 
It follows from the fact that the angles between the group and the phase velocities for the mode corresponding to the hyperbolic contours in Fig. \ref{iso1} and Fig. \ref{iso2} can be close to $\pi/2$ (the normal to the isofrequency contour shows the direction of the group velocity vector).  

At the frequencies close to the plasma frequency $\o_0$ and higher two other modes appear
with isofrequencies centered at $\Gamma$. They are shaped as two crossing ellipses. 
The modes with isofrequency curves close to $q_{y,z}=\pm k$ are still present.
The isofrequency contours for such a case (corresponding to $ka/2\pi=0.3$, $q_x=0$ and $q_x=\pi/(2a)$) are shown in Fig. \ref{iso2}. 
When $q_x$ increases at fixed frequency, the hyperbolic isofrequency contours in the plane $(q_y-q_z)$ approach the asymptotes in the same way as it happens for lower frequencies. The elliptic contours located around $\Gamma$ (see Fig. \ref{iso2}) shrink to this point when $q_x$ grows and disappear when $q_x$ becomes greater than $k_0$.

\section{Quasi-static case}

Let us consider a double wire medium in a quasi-static case when $|\vec q|a\ll \pi$ and $ka\ll \pi$.
Expanding trigonometric functions in the dispersion equation for extraordinary waves 
\r{dis2dbig} into Taylor series and keeping two first terms in those expansion, we receive to the following
equation: 
$$
(k^2-q_y^2)(k^2-q_z^2)[k^2-k_0^2(r_y,a,c)-q^2][k^2-k_0^2(r_z,a,b)-q^2]
$$
\e
=q_y^2q_z^2k_0^2(r_y,a,c)k_0^2(r_z,a,b)
\l{quasi1}
\f 
Let us consider a special case when $r_y=r_z=r$ and $a=b=c$. In that case \r{quasi1} could be simplified to the form: 
\e
\sqrt{k^2-q_y^2}\sqrt{k^2-q_z^2}(k^2-k_0^2-q_x^2-q_y^2-q_z^2) \pm q_yq_zk_0^2=0,
\l{quasi2}
\f 
where $k_0=k_0(r,a,a)$. 

We can express $q_x$ from \r{quasi2} in the form  
\e
q_x^2=k^2-k_0^2\pm {q_yq_z k_0^2\over \sqrt{k^2-q_y^2}\sqrt{k^2-q_z^2}}-q_t^2.
\l{brred}
\f

In \cite{Mario3d} the following 
approximate dispersion equation was introduced 
under the conditions $k\approx k_0$ and $q_t=\sqrt{q_x^2+q_y^2}\ll k_0$ 
(in our notations):
\e
q_x^2\approx k^2-k_0^2\pm {q_yq_z}-q_t^2.
\l{bred}\f
The birefringence of the dispersion branches near the plasma frequency 
corresponds to the two signs in the right side of \r{brred} and \r{bred}.  
The difference between equations \r{brred} and \r{bred}  is not significant if $k\approx k_0$ and $q_t=\sqrt{q_x^2+q_y^2}\ll k_0$. 

In \cite{Mario3d} the propagation of waves at the frequencies close to $\o_0$ in triple 
wire medium has been numerically studied for the case when $q_x=0$ (in-plane propagation). 
In that case the presence of $x-$wires does not influence the propagation characteristics, and our dispersion equation \r{quasi2} is applicable. 
We compare the solution of \r{quasi2} with results from \cite{Mario3d} in order to validate our
theory.

\begin{figure}[htbp]
\centering \epsfig{file=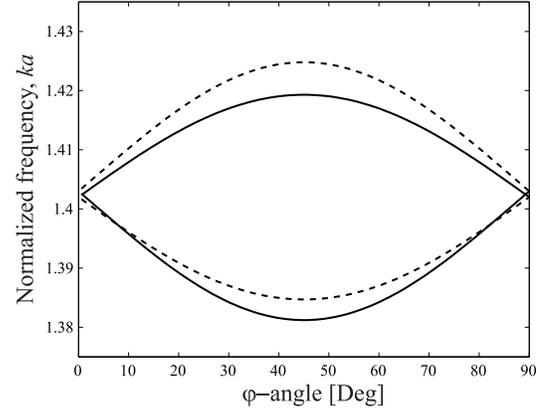, width=7cm} \caption{Dependence of the normalized wave number $ka$ on $\varphi$ angle near the "plasma" resonance of the wire medium with $r=a/100$, $q_t=0.1\pi/a$ (dashed line).
Comparison with the exact result (solid line corresponds to the
numerical data from Fig. 6, \cite{Mario3d}).} \label{f2}
\end{figure}

In \cite{Mario3d} one chose the following parameters: $r=a/100$ 
(it corresponds to $k_0a\approx 1.4$), $q_t=0.1\pi/a$. One calculates 
the medium dispersion for $q_x=0$ at the frequencies $k\approx k_0$. In Fig. 6 of this work one shows the dependence of the normalized eigenfrequency $ka$ on the angle $\varphi$. 
The angle $\varphi$ is indicated in Fig. \ref{2d}
($q_y=q_t\sin\varphi$ and  $q_z=q_t\cos\varphi$). 
The plot $ka$ vs. $\varphi$ shown in Fig. \ref{f2} represents the
comparison of \r{quasi2} with the numerical data from \cite{Mario3d}. The
upper dispersion branch corresponds to the case when before $q_yq_z$
in \r{quasi2} there is a plus sign, and the lower branch corresponds to the case when there is the minus sign.
This plot illustrates the effect of the dispersion branch birefringence near the 
"plasma frequency" of wire medium (see ellipses in Fig. \ref{iso2}). Equation \r{quasi2} shows that this effect keeps also for $q_x\ne 0$.
Fig. \ref{f2} verifies that the quasi-static equation \r{quasi2} is correct even at rather high frequencies (slightly higher than $\o_0$) outside the initial approximation 
$ka\ll \pi$.

Now, let us turn to the consideration of the effective permittivity of 2d WM.  
In the dyadic form the
tensor of effective relative permittivity of arbitrary anisotropic dielectric media can be
written as
$\=\varepsilon=\varepsilon_{xx}\vec x_0\vec x_0+\varepsilon_{yy}\vec y_0\vec y_0+\varepsilon_{zz}\vec z_0\vec z_0$. 
The dispersion equation of anisotropic dielectric (following from
Maxwell's equation and from the definition of relative permittivity in terms of
$\vec D=\varepsilon_0\=\varepsilon\cdot\vec E$) corresponds to the zero-value determinant of
the following system $(\=\varepsilon  k^2+\vec q\vec q-q^2\=I)\vec E=0$:
\e 
\left\{
\begin{array}{lcl}
(\varepsilon_{xx}k^2-q_y^2-q_z^2)E_x+q_xq_yE_y+q_xq_zE_z=0\\ 
q_xq_yE_x+(\varepsilon_{yy}k^2-q_x^2-q_z^2)E_y+q_yq_zE_z=0\\
q_xq_zE_x+q_yq_zE_y+(\varepsilon_{zz}k^2-q_x^2-q_y^2)E_z=0
\end{array}
\right.
\l{dispany}
\f 
This system helps to find the polarization of eigenmodes when $\=\varepsilon$ is known. The study of eigenmodes polarization is going to be considered in the next paper.  

The dispersion equation has the following form:
$$
(q_y^2+q_z^2-k^2\varepsilon_{xx})(q_x^2+q_z^2-k^2\varepsilon_{yy})(q_x^2+q_y^2-k^2\varepsilon_{zz})-
$$
$$
(q_y^2+q_z^2-k^2\varepsilon_{xx})q_y^2q_z^2-(q_x^2+q_z^2-k^2\varepsilon_{yy})q_x^2q_z^2-
$$
\e (q_x^2+q_y^2-k^2\varepsilon_{zz})q_x^2q_y^2-2q_x^2q_y^2q_z^2=0.
\l{f}
\f 

In \cite{Mario3d} the following expressions have been heuristically introduced for components of the permittivity of 3d WM:
$$
 \varepsilon_{xx} =1-\frac{k_0^2(r_x,b,c)}{k^2-q_x^2},\qquad  
 $$ 
 \e
 \varepsilon_{yy}=1-\frac{k_0^2(r_y,a,c)}{k^2-q_y^2},
 \quad  \varepsilon_{zz}=1-\frac{k_0^2(r_z,a,b)}{k^2-q_z^2} \l{this}. 
 \f
The effects of the low-frequency spatial dispersion are deemed to be described by terms $q_{x,y,z}$ in the denominators of the components of $\=\varepsilon$ (see also in \cite{WMPRB}).  

It has been noticed in \cite{Mario3d} that the expressions \r{this} and the dispersion equation \r{bred} fit perfectly with the results of numerical simulations for $\o\approx \o_0$. 
Following \r{this}, the components of $\=\varepsilon$ for triple WM are the permittivities of the three orthogonal simple wire media stretched along Cartesian axes. 
We have assumed that the same rule holds for 2d WM.
In case of 2d wire medium there are no $x-$directed wires and $\varepsilon_{xx}=1$. 
We have analytically verified that \r{quasi1} exactly coincides with \r{f} if the effective permittivity 
of a double WM takes the form:
 \e
 \=\varepsilon_{\rm double}=\vec x_0\vec x_0+\varepsilon_{yy}\vec y_0\vec y_0+\varepsilon_{zz}\vec z_0\vec z_0, 
 \l{pupa}\f
where $\varepsilon_{yy}$ and $\varepsilon_{zz}$ are given by the relations \r{this}. It should be noticed that the formula \r{pupa} has been obtained (very recently) for 2d WM by other authors \cite{Mariohomo} as a result of a very complicated analytical-numerical approach. 
>From \r{this} it follows, that at every point of the central isofrequency contour in Fig. \ref{iso1} (where $k<k_0$ and $q_{y,z}<k$) both components of the permittivity tensor $\varepsilon_{yy}$ and $\v_{zz}$ are negative. The propagation of such a wave ($I\ne 0$ for it and the electric field can contain $y-$ and $z-$components) is the spatial dispersion effect. 

We have also proved that the whole system \r{this} holds in our model of a triple wire media. The quasi-static analogue of \r{dis3d} has the form
$$
(k^2-q_x^2)(k^2-q_y^2)(k^2-q_z^2)[k^2-k_0^2(r_x,b,c)-q^2]\times
$$
$$
[k^2-k_0^2(r_y,a,c)-q^2][k^2-k_0^2(r_z,a,b)-q^2]
$$
$$
-(k^2-q_x^2)[k^2-k_0^2(r_x,b,c)-q^2]q_y^2q_z^2k_0^2(r_y,a,c)k_0^2(r_z,a,b)
$$
$$
-(k^2-q_y^2)[k^2-k_0^2(r_y,a,c)-q^2]q_x^2q_z^2k_0^2(r_x,b,c)k_0^2(r_z,a,b)
$$
$$
-(k^2-q_z^2)[k^2-k_0^2(r_z,a,b)-q^2]q_x^2q_y^2k_0^2(r_x,b,c)k_0^2(r_y,a,c)
$$
\e
+2q_x^2q_y^2q_z^2k_0^2(r_x,b,c)k_0^2(r_y,a,c)k_0^2(r_z,a,b)=0.
\l{quasi3}
\f 
It coincides with the dispersion equation \r{f} if the permittivity takes the form (see relations \r{this}):
\e
\=\varepsilon_{\rm triple}=\varepsilon_{xx}\vec x_0\vec x_0+\varepsilon_{yy}\vec y_0\vec y_0+\varepsilon_{zz}\vec z_0\vec z_0.
\l{pupa3d}
\f

We have thus verified, that dielectric permittivities for 2d and 3d wire media in the form \r{pupa} and \r{pupa3d} suggested in \cite{Mario3d} and \cite{Mariohomo} fit successfully in our theory.
We have analytically verified, that the quasi-static analogues of dispersion equations \r{dis2d} and \r{dis3d} in the form \r{quasi2} and \r{quasi3} coincide with the dispersion equations of anisotropic dielectrics \r{pupa} and \r{pupa3d}.

\section{Conclusion}
In the present paper we have generalized a recently developed analytical theory of a simple wire media to the case
of double wire media and obtained some results for triple WM. We have validated our theory by comparison with \cite{Mario3d} and proved that the effective permittivity of 2d and 3d WM introduced in \cite{Mario3d} and \cite{Mariohomo} fits our dispersion equations fairly well. 
 
We have theoretically revealed the effects of low-frequency spatial dispersion for 2d WM such as:
\begin{itemize}
\item Propagation of $z-$polarized TLM along $y-$wires is not suppressed by the presence of $z-$wires (the same is correct for the $y-$polarized TLM propagating along $z$).
\item There are TLM which can exist in both $y-$ and $z-$arrays simultaneously. These modes do not transport energy, since the directions of the currents in wires are alternating along the $x-$axis.  
\item There are two propagating modes 
at low frequencies $\o<\o_0$ which are not TLM and not ordinary waves. 
One mode has non-zero electric field component in the plane $(y-z)$ whereas both $y-$ and $z-$components of the permittivity tensor are negative.
For the other one the isofrequency contour is nearly hyperbolic. 
\item Near the plasma frequency the two other waves appear with crossing 
isofrequency contours. 
\end{itemize}
The materials under consideration could find various applications due to the properties discussed in this paper.
We would like to note especially such applications as creation of low frequency super-prism and design of materials with negative refraction. Those properties of double WM will be discussed in the next paper.

Finally, let us discuss the problem of the homogenization of WM. 
The equation \r{f} relates three unknown components of $\=\varepsilon$, three components of the wave vector $\vec q$ and the frequency (or wave number $k$). The components of $\vec q$ are related through dispersion equation with $k$. It is clear that the problem of $\=\v$ has no unique solution in thus formulation. The same concerns 2d WM. Though \r{this} fits our dispersion equations this result is heuristic and the permittivity has been introduced and not derived. Is it reasonable to try to find other possible expressions for $\=\epsilon$? 

It is well-known that the effective material parameters of spatially dispersive media have another meaning that those of continuous media. The effective susceptibility of such media in presence of a point source depends on the source position and has nothing to do with the effective medium susceptibility for plane waves. The usual boundary conditions are not valid on the medium interface. So, the material parameters are not very helpful for solving the boundary problem for media with spatial dispersion. The goal of the homogenization of WM is modest: to describe the low-frequency propagating properties of an infinite medium in terms of those parameters. Therefore all we need is to introduce the permittivity which would 1) describe all effects we can reveal solving the correct (quasi-static) dispersion equation and 2) allow to find the polarization of eigenmodes correctly.
For both 2d and 3d WM the permittivity \r{this} comprises all the dispersion properties at low frequencies ($ka <1$). As to eigenwaves polarization, the result \r{this} requests further studies which will be also presented in the next paper.

\section{Acknowledgement}

Authors are grateful to Mario Silveirinha for very fruitful discussion.

\bibliography{2dwmPRE3}

\end{document}